\documentclass[a4paper,11pt]{article}
\usepackage{pos}
\title{Connecting Inflation and Low Energy Phenomenology in an Extended Two-Higgs-Doublet-Model
}
\ShortTitle{Connecting Inflation and Low Energy Phenomenology in 2hdSMASH}
\author*[a]{Juhi Dutta}
 
\affiliation[a]{Homer L. Dodge Department of Physics and Astronomy,
University of Oklahoma,\\ Norman, OK 73019, USA}
 
\emailAdd{juhi.dutta@ou.edu}
 
 \abstract{ We investigate the Two-Higgs-Doublet-Standard Model-Axion-Seesaw-Higgs-Portal inflation (2hdSMASH) model consisting of two Higgs doublets, a Standard Model (SM) singlet complex scalar and three SM singlet right-handed Majorana neutrinos that can address the strong CP problem, dark matter, neutrino masses, baryogenesis and inflation. We identify inflationary directions consistent with successful inflation and theoretical constraints including perturbative unitarity, boundedness-from-below conditions and high scale validity upto the PLANCK scale. 
   Further,  we present representative benchmark points satisfying theoretical and experimental constraints potentially accessible at  future colliders.}  
\FullConference{The European Physical Society Conference on High Energy Physics (EPS-HEP2023)\\
 21-25 August 2023\\
Hamburg, Germany\\}

\begin{document}
\maketitle

\section{Introduction}
Despite the tremendous success of the Standard Model (SM), one is compelled to venture beyond in order to understand several longstanding problems of Nature such as the presence of dark matter, non-zero neutrino masses,  strong CP problem, baryogenesis and explaining inflation, a prime candidate to understand the universe before the Big-Bang nucleosynthesis. 
While several beyond Standard Models (BSM)  address one or more aspects of these issues, a minimal extension of SM, including three right-handed Majorana neutrinos, an exotic quark along with a complex singlet scalar and a KFSZ like axion model, SMASH~\cite{Ballesteros:2016xej},  represents a   complete model of particle physics and cosmology by addressing all the issues of Nature.  Possible non-minimal extensions include the Two-Higgs-Doublet-Standard Model-Axion-Seesaw-Higgs-Portal inflation (2hdSMASH) model~\cite{Matlis:2022iwr,Dutta:2023lbw} consisting of two Higgs doublets, a Standard Model (SM) singlet complex scalar and three SM singlet right-handed neutrinos.  The field content of the 2hdSMASH model in order to explain neutrino masses, baryogenesis, the strong CP problem, and dark matter was introduced in $\nu$DFSZ~\cite{Clarke:2015bea} model where the authors discuss the high scale validity and the technical naturalness of the model while Ref.~\cite{Espriu:2015mfa} looks into the low scale phenomenology of the Higgs sector.  
Other variations such as VISH$\nu$~\cite{Sopov:2022bog} address the inflationary directions for the extended  $\nu$DFSZ with a trilinear U(1) breaking term. 

In this work, we consider the interplay of inflation and low energy phenomenology in 2hDSMASH (with a quartic U(1) breaking term) consistent with all theoretical and experimental constraints and valid all the way upto the PLANCK scale. Finally, we present some benchmarks that are allowed by current data and within the reach of the HL-LHC.

\section{The Model}
The 2hdSMASH model consists of the Type II 2HDM with two Higgs doublets, $\Phi_1, \Phi_2$ extended with a complex scalar singlet $S$ and three right-handed Majorana neutrinos $N_R$ under  a global U(1) Peccei-Quinn symmetry. The charges of the fields are shown in Table~\ref{TabPQ}.
\begin{table}[h]
\small
\begin{center}
 \begin{tabular}{|c|c|c|c|c|c|c|c|c|c|}
   \hline 
  Field & $S$ & $\Phi_1$ & $\Phi_2$ & $q_L$ & $u_R$ & $d_R$ & $l_L$ & $N_R$ & $e_R$ \\
   \hline
  Charge & $X_S$ & $X_1$ & $X_2$ & $X_q$ & $X_u$ & $X_d$ & $X_l$ & $X_N$ & $X_e$ \\
   \hline
Value & $1$ & $-\frac{2x}{x-1}$ & $\frac{2}{1-x}$ & $0$ & $\frac{2}{1-x}$ & $\frac{2x}{x-1}$ & $\frac{3}{2}-\frac{2}{1-x}$ & -$\frac{1}{2}$ & $\frac{7}{2} -\frac{4}{1-x}$ \\
   \hline 
 \end{tabular}
\end{center}
 \caption{The charges of fields under the PQ symmetry~\cite{Clarke:2015bea}. Here $x\neq 1$.
}
 \label{TabPQ}
\end{table}  
The most general renormalizable scalar potential is 
  \begin{align*}
  \scriptsize
 V(\Phi_1,\Phi_2,S) =&\,
  M^2_{11}\, \Phi_1^\dagger \Phi_1
  + M^2_{22}\, \Phi_2^\dagger \Phi_2 
  +M_{SS}^2\, S^*S
  + \frac{\lambda_1}{2} \left( \Phi_1^\dagger \Phi_1 \right)^2
  + \frac{\lambda_2}{2} \left( \Phi_2^\dagger \Phi_2 \right)^2
  + \frac{\lambda_S}{2}\left(S^*S\right)^2
  \nonumber\\ 
&
  + \lambda_3\, \left(\Phi_1^\dagger \Phi_1\right) \left(\Phi_2^\dagger \Phi_2\right)
  + \lambda_4\, \left(\Phi_1^\dagger \Phi_2\right) \left(\Phi_2^\dagger \Phi_1\right)
+\lambda_{1S}\left(\Phi_1^\dagger\Phi_1\right)\left(S^*S\right)
        +\lambda_{2S}\left(\Phi_2^\dagger\Phi_2\right)\left(S^*S\right)\nonumber \\
        & -\lambda_{12S}\left(\Phi_2^\dagger\Phi_1 S^2+h.c.\right)\,,
\label{EqnuDFSZ}
\end{align*}
and the Yukawa Lagrangian is
\begin{equation*}
\scriptsize
  -\mathcal{L}_Y =  Y_u\overline{q_L}\tilde\Phi_2 u_R + Y_d\overline{q_L}\Phi_1 d_R 
    + Y_e \overline{l_L}\Phi_1 e_R + Y_\nu \overline{l_L}\tilde\Phi_1 N_R 
    + \frac12 y_N \overline{(N_R)^c}S N_R + {\rm h.c.}\,,
\label{EqYuk}
\end{equation*}
where, 
 \begin{equation*}
\Phi_i = 
\begin{pmatrix}
  \phi^{\pm}_i  \\
  \frac{1}{\sqrt{2}}(v_i+H_i+ i A_i)\\
 \end{pmatrix}, \,\,
  S = \frac{1}{\sqrt{2}}(v_S+ H_S + i A_S)
\end{equation*}
  $v (=\sqrt{v^2_1+v^2_2})$ and $v_S$ are the electroweak and singlet vacuum expectation values respectively and $v_i$ ($i=1,2$) refer to the vacuum expectation values obtained by the neutral scalars of the  Higgs doublets  $\Phi_i$ and $M_{11}, M_{22}, \lambda_1-\lambda_4$ are the usual parameters of 2HDM while $M_R$, $M_{SS}$, $\lambda_S$, $\lambda_{1S}$, $\lambda_{2S}$ and $\lambda_{12S}$ are the mass matrix of the right-handed neutrino, mass and self-coupling of the complex scalar singlet and the portal couplings with the Higgs doublets respectively with $\lambda_{12S}$ being the $U(1)$  symmetry breaking term. 

After electroweak-symmetry breaking, the Higgs sector consists of a pair of charged Higgses $H^{\pm}$, two CP-odd states $A$ and $a$ (axion), and three neutral CP-even states, $h$, $H$, and $s$.  Owing to the constraints on $v_S$ from the axion decay constant and from satisfying relic abundance, the large hierarchy between $v$ and $v_S$ leads to a hierarchical scalar sector spectrum with a light axion with mass$\propto \frac{1}{v_S}$,  the lightest CP-even Higgs mass $\propto v$ and the heavy CP-even scalar $H$, $s$ and charged Higgs $H^{\pm}$ masses $\propto v_S$. 

\section{Results}

In this section, we  discuss the inflationary directions and constraints on the inflationary observables and present representative benchmarks which satisfy all theoretical conditions such as the boundedness-from-below conditions, perturbative unitarity, high scale validity and experimental constraints from inflation, dark matter relic abundance and collider searches.  For axionic dark matter, the present constraints on axion decay constant, $f_a$, and thermal relic abundance require $v_S$ to be $\mathcal{O}(10^9 -10^{10})$ GeV.  

\begin{figure}
    \centering
    \includegraphics[scale=0.5]{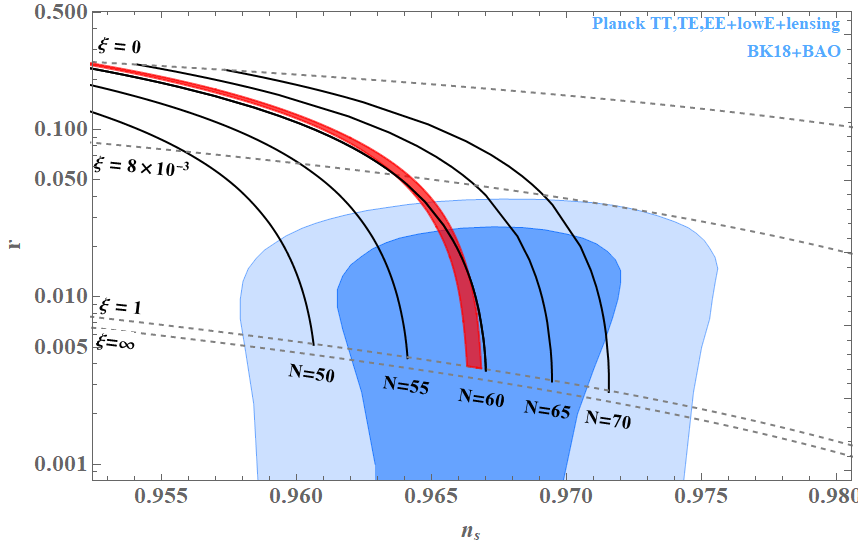}
    \caption{Variation of $n_s$ vs. $r$ for different choices of non-minimal couplings $\xi_S$ constrained by PLANCK/BICEP data.  The red line denotes the inflationary predictions of  2hDSMASH and the blue coloured regions indicate the 95$\%$ C.L. and 68$\%$ C.L. allowed regions from PLANCK  data.}
    \label{fig:nsr}
\end{figure}

Our focus is on the hierarchical regime $\xi_S \gg \xi_1, \xi_2$ where $\xi_i$ ( $i$ = 1, 2, $S$ ) are the non-minimal couplings of the Higgs fields and the singlet scalar field to the Ricci scalar with 
inflation  along the  singlet directions (PQI), or mixed singlet-Higgses directions (PQTHI). 
In Table~\ref{tab:inflation_cond} we summarize the inflationary conditions and  Einstein frame slow-roll inflationary potential for the different inflationary directions.  
In order to preserve perturbative unitarity and obeying
the PLANCK 2018 constraint on the scalar perturbation amplitude $A_{S}\sim 10^{-10}$, requires  $\tilde{\lambda}\sim 10^{-10}\xi_{S}^{2}$ where 
$\tilde{\lambda}$ is the effective quartic coupling of the singlet scalar of the inflationary potential in the Einstein frame. This sets a constraint $\tilde{\lambda}\sim 10^{-10}$ for $\xi_S\sim\mathcal{O}(1)$  where $\tilde{\lambda_S} = \lambda_{j}$ for $j = s,sh_1,sh_2,sh_{12}$ for the various inflationary directions as seen in Table~\ref{tab:inflation_cond}. The effective quartic couplings for the different inflationary directions are  $\lambda_{sh_i}=\lambda_S-\frac{\lambda^2_{iS}}{\lambda_i}$ and 
$\lambda_{sh_{12}}=\lambda_S-\frac{\lambda_1\lambda^2_{2S}+\lambda_2\lambda^2_{1S}-2\lambda_{1S}\lambda_{2S}\lambda_{34}}{\lambda_{1}\lambda_2-\lambda_{34}}$ where ($i=1,2$ and $\lambda_{34}=\lambda_{3}+\lambda_{4}$). For $\xi_S \leq 1$, \begin{align}
&8.5\times 10^{-3}\lesssim\xi_S\lesssim 1~~\text{implying}~~9\times 10^{-10}\gtrsim \tilde{\lambda}\gtrsim 4.5\times 10^{-13}\,.
\end{align}
Other inflationary observables, such as the spectral tilt $n_{s}$ and the tensor-to-scalar ratio $r$, computed for 2hdSMASH are constrained from the PLANCK/BICEP data for different  values of the non-minimal couplings of the singlet are as seen in Fig.~\ref{fig:nsr}. 
In order to ensure the validity of the model upto the PLANCK scale, one must ensure that the couplings run up to the PLANCK scale satisfying all theoretical constraints such as boundedness-from-below, perturbative unitarity and inflationary conditions. 
Further to ensure $\tilde{\lambda}_{S}\sim 10^{-10}$, at all scales sets an upper limit of $\mathcal{O}$(10$^{-5}$) for the portal couplings as well as the neutrino Yukawa couplings~\cite{Matlis:2022iwr,Dutta:2023lbw}. Such tiny couplings favour a technically natural theory as first pointed out by ~\cite{Clarke:2015bea}. Further constraints on the right-handed neutrino masses and Yukawa couplings arise from baryogenesis and neutrino oscillations~\cite{Matlis:2022iwr,Dutta:2023lbw}. Vacuum stability of the model also constrains the 2HDM couplings, and particularly $\lambda_2$ and $\lambda_{34}$ ( = $\lambda_3+\lambda_4$) to ensure the scalar potential remains stable up to PLANCK scale~\cite{Matlis:2022iwr,Dutta:2023lbw}. 
\begin{table}[htb!]
\scriptsize
\begin{center}
\begin{tabular}{ |c|c|c| }
 \hline
 Inflationary direction &  Inflationary conditions & Einstein frame slow roll potential  \\
 \hline \hline
$s\,h_{1}$&     \begin{tabular}{@{}c@{}}\\ $\lambda_{2S}\lambda_1-\lambda_{1S}\lambda_{34}\geq 0$~,~ $\lambda_{1S}\lambda_2-\lambda_{2S}\lambda_{34}\leq 0$\\~\\
$\lambda_{1S}\leq 0$~,~ $\lambda_{2S}\geq 0$\\~\end{tabular}
& $\dfrac{\lambda_{sh_1}}{8}s^4\left( 1 + \xi_S\frac{s^2}{M_P^2}\right)^{-2}$ \\ 
\hline 
$s\, h_{2}$&    \begin{tabular}{@{}c@{}}\\
$\lambda_{2S}\lambda_1-\lambda_{1S}\lambda_{34}\leq 0$~,~ $\lambda_{1S}\lambda_2-\lambda_{2S}\lambda_{34}\geq 0$\\~\\
$\lambda_{1S}\geq 0$~,~ $\lambda_{2S}\leq 0$\\~\end{tabular}& $\dfrac{\lambda_{sh_2}}{8}s^4\left( 1 + \xi_S\frac{s^2}{M_P^2}\right)^{-2}$ \\ 
\hline 
$s\,h_{12}$&  \begin{tabular}{@{}c@{}}\\ $\lambda_{2S}\lambda_1-\lambda_{1S}\lambda_{34}\leq 0$~,~ $\lambda_{1S}\lambda_2-\lambda_{2S}\lambda_{34}\leq 0$\\~\\
$\lambda_{1S}\leq 0$~,~ $\lambda_{2S}\leq 0$\\~
   \end{tabular}
& \begin{tabular}{@{}c@{}}\\
 $\dfrac{\lambda_{sh_{12}}}{8}s^4\left( 1 +\xi_{2}\frac{s^2}{M_P^2}\right)^{-2}$ \\~
\end{tabular}\\
\hline
$s$ 
&  \begin{tabular}{@{}c@{}}\\ $\lambda_{2S}\lambda_1-\lambda_{1S}\lambda_{34}\geq 0$~,~ $\lambda_{1S}\lambda_2-\lambda_{2S}\lambda_{34}\geq 0$\\~\\
$\vee ~\lambda_{1S,2S}\ll \lambda_{S}$\\~
   \end{tabular}
& \begin{tabular}{@{}c@{}}\\
 $\dfrac{\lambda_{S}}{8}s^4\left( 1 +\xi_{S}\frac{s^2}{M_{p}^2}\right)^{-2}$ \\~
\end{tabular}\\
\hline  
\end{tabular}
\end{center}
\caption{Conditions and characteristics for PQI and PQTHI, i.e. $s$- and $s h_{1,2,12}$-inflation, with $\xi_{S}\gg\xi_{1,2}$~\cite{Dutta:2023lbw} where $\xi_{1,2,S}$ are the non-minimal couplings of the Higgs doublets and  singlet scalar field respectively. }
\label{tab:inflation_cond}
\end{table}
Taking into account all these constraints, we now present some representative benchmarks consistent with all theoretical and experimental constraints in Table~\ref{tab:bp}. We observed that the favoured range of portal couplings and $v_S$, and $\tan \beta$ have a strong bearing on the mass of the heavy Higgs spectra. Owing to the large hierarchy between $v$ and $v_S$ and coupled with the tiny portal couplings along with $\lambda_S$, the mass spectra consists of a 125 GeV SM-like Higgs, a heavy scalar $\mathcal{O}(v_S)$, while the other CP-even Higgs, CP-odd Higgs and charged Higgses may lie around the TeV scale  depending on the values of the portal couplings $\lambda_{12S},\lambda_{1S},\lambda_{2S}$ and the singlet self coupling $\lambda_S$ as seen in \textbf{BP1}-\textbf{BP4}. 
\begin{table}[ht]
\scriptsize
\begin{center}
 \begin{tabular}{|c|c|c|c|c|}
  \hline
  Parameters &\textbf{BP1}&\textbf{BP2} &\textbf{BP3} & \textbf{BP4} \\
  \hline
  $\lambda_1$ &0.07 &0.07 & 0.07&0.07  \\
  $\lambda_2$ & 0.263&0.257&0.258& 0.257 \\
  $\lambda_3$ &0.60 &0.24 & 0.54 & 0.24  \\
  $\lambda_4$ &-0.4&0.27&-0.14 &-0.28 \\
  $\lambda_S$ &6.5$\times 10^{-10}$&1.0$\times 10^{-10}$ &1.0$\times 10^{-10}$& 1.0$\times 10^{-10}$   \\
  $\lambda_{1S}$ &-6.59$\times 10^{-6}$&4.8$\times 10^{-14}$&4.8$\times 10^{-14}$ &3.6$\times 10^{-13}$ \\
  $\lambda_{2S}$ &1.0$\times 10^{-15}$&1.0$\times 10^{-15}$&1.0$\times 10^{-15}$ &1.0$\times 10^{-15}$   \\
  $\lambda_{12S}$ &2.5$\times 10^{-16}$&2.5$\times 10^{-16}$& 2.5$\times 10^{-16}$&2.5$\times 10^{-16}$  \\
  $\tan \beta$  &5.5&26&26 &18   \\
$Y_{N,1}$& $9\times 10^{-4}$&$4\times 10^{-5}$&$4\times 10^{-5}$ & $10^{-4}$\\
    $Y_{\nu,3}$& $5.175\times 10^{-3}$&$1.09\times 10^{-3}$ &$1.09\times 10^{-3}$&$1.2\times 10^{-3}$\\ 
  $v_S$ & 3.0$\times 10^{10}$&3.0$\times 10^{10}$&3.0$\times 10^{10}$ &3.0$\times 10^{10}$   \\
  \hline
  $m_h\left(\text{GeV}\right)$  &125.3  &  125.0   & 124.9 & 124.4  
 \\
  $m_H\left(\text{GeV}\right)$  &799.4&1711.5&1711.5 &1425.2 \\
  $m_{s}\left(\text{GeV}\right)$  &6.7$\times 10^{5}$&3.0$\times 10^{5}$&3.0$\times 10^{5}$ &3.0$\times 10^{5}$ \\
  $m_A\left(\text{GeV}\right)$  &799.5&1711.5&1711.5 &1425.2 \\
  $m_{H^{\pm}}\left(\text{GeV}\right)$  &807.0&1709.1&1712.8 &1422.2 \\ 
  \hline
 \end{tabular}
 \caption{List of benchmarks passing the theoretical constraints and experimental constraints.}
 \label{tab:bp}
\end{center}
\end{table} 

\begin{figure}[ht]
    \centering
     \includegraphics[scale=0.32]{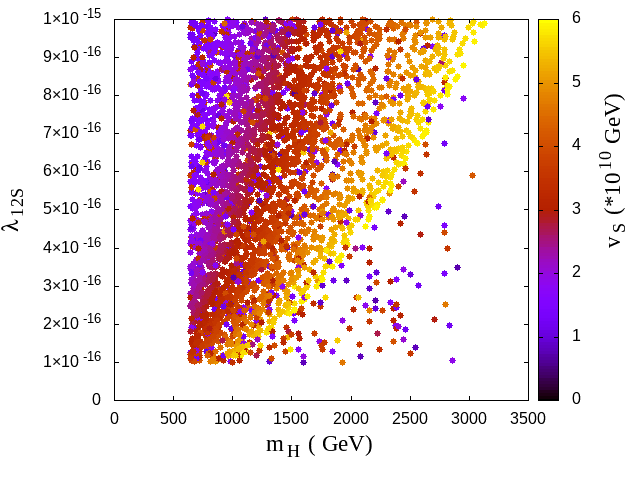} 
     \includegraphics[scale=0.32]{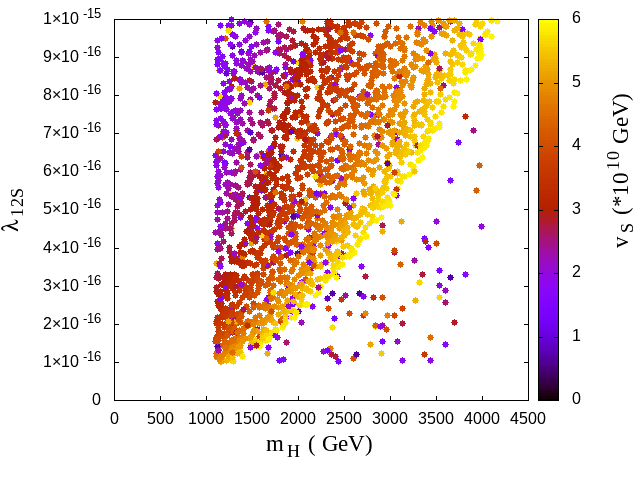}
         \caption{Variation of $\lambda_{12S}$ vs. $m_H$ for $\tan \beta =5.5, 10$. }
    \label{fig:l12smH}
\end{figure}

 For fixed 2HDM couplings and portal couplings, $\lambda_{1S}$ and $\lambda_{12S}$, the spectrum is completely determined by $\tan \beta, \lambda_{12S}$ and $v_S$.  We illustrate this by scanning over the allowed range of $v_S$ and varying $\lambda_{12S}\sim 10^{-16}-10^{-15}$ keeping the other parameters fixed as in \textbf{BP1}. 
Fig.~\ref{fig:l12smH} shows the variation of $\lambda_{12S}-m_{H}$ plane where $m_H$ is the mass of the heavy CP-even Higgs over the allowed range of $v_S$ and $\lambda_{12S}\sim 10^{-16}-10^{-15}$, allowed by the constraints from Higgs physics, for two values of $\tan \beta = 5.5$ and 10. In Fig.~\ref{fig:l12smH} (left) , for a fixed $v_S$, with decreasing $\lambda_{12S}$, $m_H$ decreases and for a fixed value of $\lambda_{12S}$, $m_H$ increases with increase in $v_S$. Further, one observes an overall increase in the mass scale with an increase in $\tan \beta$ as seen in Fig.~\ref{fig:l12smH} (right).  Thus, over the entire allowed range of $v_S$ there is ample parameter space allowed by current experimental data allowing $m_H$ within a few TeV, and within reach of the HL-LHC. 
 
\section{Summary and Conclusions}
 We have investigated the 2hdSMASH model consisting of the Type II Two Higgs Doublet model augmented with a complex singlet scalar and three right-handed Majorana neutrinos. We have considered the impact of all theoretical constraints including boundedness-from-below, perturbative unitarity, and high-scale validity and experimental  constraints from inflation, dark matter  relic abundance and from colliders are taken into account to obtain a viable parameter space for the model. 
  For the singlet and singlet-Higgs inflationary directions, the effective coupling of the inflationary potential $\tilde{\lambda}$ is constrained to be $\mathcal{O}(10^{-10})$ from the CMB data. Further constraints on the singlet portal couplings ($\sim 10^{-5}$) arise from the RGE running of the couplings to preserve high scale validity upto the PLANCK scale.  In addition, constraints from baryogenesis (via leptogenesis) and neutrino oscillations are also accommodated which further  constrain the neutrino Yukawa couplings.  The large hierarchy of $v_S\gg v$ and tiny portal couplings leads to a naturally compressed spectra for a heavy CP-even Higgs, CP-odd Higgs and pair of charged Higgses $H^{\pm}$ with masses near the $\mathcal{O}$(TeV) scale. Some representative benchmarks are presented with a CP-even heavy Higgs scalar, CP-odd pseudoscalar and pair of charged Higgses with masses $\sim \mathcal{O}$ (TeV) are  within the reach of HL-LHC for future phenomenological studies. 
\section*{Acknowledgments}
The author acknowledges support from the  HEP  Dodge Family Endowment Fellowship at the Homer L.Dodge Department of Physics $\&$ Astronomy at the University of Oklahoma. The author thanks her collaborators  M.Matlis, G.Moortgat-Pick and A.Ringwald for successful completion of the work, Ref~\cite{Dutta:2023lbw}, on which this talk is based on.

\end{document}